# Extemporaneous Mechanochemistry: Shockwave Induced Ultrafast Chemical Reactions Due to Intramolecular Strain Energy

Brenden W. Hamilton[1], Matthew P. Kroonblawd[2], Alejandro Strachan[1]*

[1]School of Materials Engineering and Birck Nanotechnology Center, Purdue University, West Lafayette, Indiana, 47907 USA
[2]Physical and Life Sciences Directorate, Lawrence Livermore National Laboratory, Livermore, California 94550, United States

## Abstract

Regions of energy localization referred to as hotspots are known to govern shock initiation and the run-to-detonation in energetic materials. Mounting computational evidence points to accelerated chemistry in hotspots from large intramolecular strains induced via the interactions between the shockwave and microstructure. However, definite evidence mapping intramolecular strain to accelerated or altered chemical reactions has so far been elusive. From a large-scale reactive molecular dynamics simulation of the energetic material TATB, we map molecular temperature and intramolecular strain energy prior to reaction to decomposition kinetics. Both temperature and intramolecular strain are shown to accelerate chemical kinetics. A detailed analysis of the atomistic trajectory shows that intramolecular strain can induce a mechanochemical alteration of decomposition mechanisms. The results in this paper can inform continuum-level chemistry models to account for a wide range of mechanochemical effects.

\* strachan@purdue.edu

Shock loading can rapidly initiate numerous phenomena such as phase transitions[1,2] and chemistry[3–5], as well as plasticity, fracture, and failure[6,7]. While these processes are also observed under static and slower loading rates[8–11], the ultrafast loading and transient states induced by a shockwave can result in exotic processes such as virtual melting[12,13], metallization of molecular crystals[14,15], and mechnochemistry[16–18]. A growing body of evidence suggests that ultrafast non-equilibrium processes can play a significant role in the initiation of high explosives (HEs).

The initiation of chemical reactions and transition to detonation in composite HE formulations under dynamical loading is dominated by energy localization into hotspots via the interaction of the shockwave with the microstructure[19]. These hotspots are almost exclusively characterized by their temperature fields[20–23]. However, recent molecular dynamics (MD) simulations indicate a deviation from this traditional picture. Dynamical hotspots can be significantly more reactive than thermodynamically equivalent ones created under non-shock conditions.[24,25] Path-dependent reactivity and mechanochemistry have been proposed to explain these observations. However, direct isolation of the underlying non-equilibrium mechanochemical processes and their effect on shock-induced chemistry remains elusive.

In the context of gas-phase chemistry, quantum chemistry[26] studies have revealed how individual molecules respond to external forces[27,28] and can result in alternate or forbidden reaction pathways[29,30]. Applications-oriented studies have revealed that mechanochemical processes can couple synergistically with environmental factors to accelerate polymer degradation[31,32], optimize industrial reactions[33,34], and attenuate shock waves[35,36].

Mechanochemistry has also been studied in condensed phases. Examples include pure carbon[37], polypeptides,[38] and HE crystals[39]. These studies show a variety of non-thermal reaction acceleration mechanisms and the potential for altered reaction pathways that arise from deformation mechanisms that operate on multi-nanometer length scales well beyond that of the individual molecules. These state-of-the-art studies for condensed matter mechanochemistry deliberatively introduce mechanical perturbations to affect chemistry. How possible mechanochemistry occurs spontaneously under realistic loading conditions and how the various chemical paths conspire to determine the global response of the system remain open questions.

Most of the above work has focused on how to activate and control mechanochemistry, henceforth denoted as "premeditated mechanochemistry". These are cases in which the system or experiment is designed to undergo mechanochemical activation when triggered, for both basic sciences and engineering materials purposes, as the inherent premeditated nature of the event makes it easier to study and harness. We instead focus on the distribution of molecular scale responses originating from realistic dynamical loading conditions, i.e., "extemporaneous mechanochemistry". Examples of this can include impacts of planetary bodies[4,38,40], detonation events in solid energetics[15,16], and shearing of materials in earth's core and between tectonic plates[37,41]. The key to understanding extemporaneous mechanochemistry is not just the trends, but to map the physical chemistry processes involved to understand the overall governing dynamics of mechanochemistry. Here, we aim to study the entire extemporaneous event via the shock initiation of a porous sample of TATB (1,3,5-triamino-2,4,6-trinitrobenzene), an insensitive HE material. Our simulation set up is designed with inspiration from previous studies[24,42,43] to provide the best opportunity to see an extemporaneous mechanochemical event.

Recent MD simulations revealed that the shock-induced collapse of porosity results not just in energy localization in the form of heating but also in significant intramolecular strains[42,43]. The excess potential energy (PE), or intramolecular strain energy[44], is readily stored in modes relevant to prompt chemical initiation and exceeds the expected value based on the thermal vibrational energy[42]. Similarly, quantum-chemistry calculations indicate that HOMO-LUMO gap closure arises in TATB when nitro groups are deformed out of plane, correlating mechanochemical effects with torsional (4-body) strains[45]. We define the excess, or latent, PE for each molecule as the difference between the actual internal PE ($U_{intra}$) and that expected from equipartition of energy corresponding to the molecule's instantaneous temperature:

$$U_{latent} = [U_{intra} - U_0] - \frac{3N-6}{2} k_b (T - 300)$$

where $U_{intra}$ is determined from a single point calculation of each individual molecule only considering internal interactions, $U_0$ is the average $U_{intra}$ in a 300 K perfect crystal, N is the number of atoms in the molecule, and T is the roto-vibrational kinetic energy of the molecule in units of kelvin. Additionally, we also quantified the strain energy originating from the ReaxFF four-body terms[46] (torsional and improper dihedrals) which will be referred to as $U_{4-body}$.

We performed large-scale, *reactive* MD simulations to characterize the effect of the latent energy in hotspots to induce extemporaneous mechanochemistry. Using a cylindrical pore of 40nm in diameter, we focus on relatively strong shocks where the resulting hotspots can transition to a deflagration wave within 10s of picoseconds[24,47]. Figure 1 shows the evolution of temperature and $U_{4-body}$ in the central hotspot, radiating shear bands, and surrounding material. Directly after pore collapse, both values rise significantly within the hotspot area. Following this initial formation stage, the high temperature region grows as the hotspot transitions towards a deflagration and the reaction zone expands driven by exothermic reactions. In contrast, the $U_{4-body}$ term begins to relax within a few picoseconds of the collapse. This is driven by chemical reactions, which alleviate molecular strain on the 4-body terms measured here. Additionally, regions that react without significant molecular strain will also tend towards $U_{4-body}$ values of zero as most gaseous products do not have four-body intramolecular components ($CO_2$, $H_2O$, $N_2$, etc.).

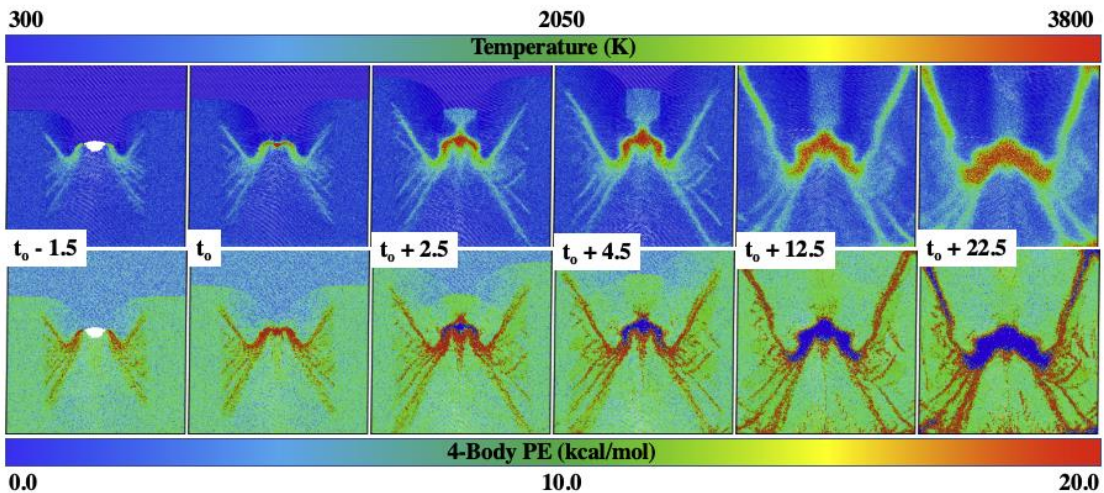

*Figure 1: Time history of roto-vibrational temperature and 4-body PE in a reacting pore-collapse hot spot in TATB. The time origin $t_o$ corresponds to complete collapse of the pore, with time offsets given in picoseconds.*

To characterize the role of $U_{latent}$ on chemical kinetics we characterize the thermodynamic state of each molecule that fully reacts during the simulation time by computing its temperature and $U_{Latent}$ right before reaction initiation, at time $t = t_1 - 0.1\ ps$. Here, $t_1$ is the initial decomposition time, defined as the first reaction in the molecule. To obtain a measure of total reaction time, a second time (t₂) is defined as the first C-C ring bond scission. We define reaction time as $\tau_{Rxn} = t_2 - t_1$. We cluster all reacting molecules by their $U_{latent}$ and temperature using an unsupervised k-means cluster analysis, see Figure 2a. Figure 2b shows the resulting cluster centroids for 6 clusters, mapped in temperature-strain energy space. All points in Figure 2 are colored by their cluster's average $\tau_{Rxn}$, also indicated as inset text. Supplemental Materials section SM-1 lists the coordinates and reaction times for the cluster centroids. While clustered molecules are localized in $T$-$U_{latent}$ space, molecules within a given cluster are spatially distributed about the hotspot (see Figure 2c).

From the cluster centroids, two distinct trends are clear. The first, which is anticipated from Arrhenius kinetics, is that increasing temperature leads to faster reactions. Clusters 1, 2, and 3 all have similar intramolecular strain energy and exhibit a systematic decrease in $\tau_{Rxn}$ with increasing temperature. The second trend, which is somewhat expected from premeditated mechanochemistry studies, is a decrease in $\tau_{Rxn}$ with an increase in $U_{Latent}$. Clusters 2, 4, and 5 have roughly the same average temperature and exhibit a reduction in reaction time with increasing strain. Note that Cluster 5 has a significantly lower $\tau_{Rxn}$ than Cluster 3, despite being 240 K colder than Cluster 3.

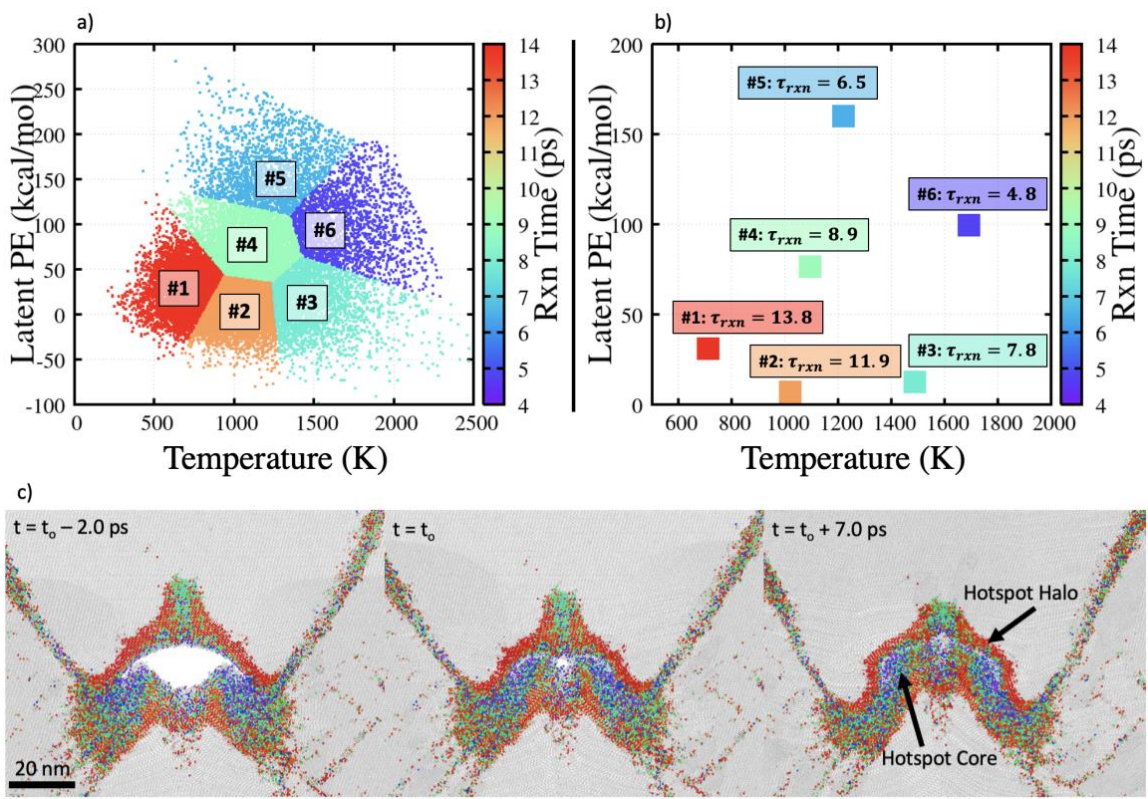

Figure 2: Molecular descriptors (a) and cluster centroids (b) colored by $\tau_{Rxn}$ plotted in $T$-$U_{latent}$ space. (c) Spatial mapping of clustered molecules about the hotspot during and after its formation. Molecules are colored by cluster ID following the convention in panels a and b, with non-clustered molecules colored white.

Spatial mapping of clustered molecules reveals several important trends. First, the slowest reacting clusters (i.e., 1 and 2) are mostly found in a "halo" around the "core" of the hotspot. In contrast, the hotspot core contains a mixture of molecules from the mechanochemically activated clusters (i.e., 4, 5, and 6) and the cluster of hot undeformed molecules (Cluster 3). This indicates that the accelerated reaction kinetics observed within hotspots determined from previous reactive MD simulations of pore collapse[24] is almost certainly mechanochemical in origin. There is also a noticeable (but smaller) population of cluster 4, 5, and 6 molecules within the shear bands that emanate from the central hotspot, which is consistent with previous work[17] showing accelerated reaction kinetics in shear bands. It is evident that the molecules with the largest $U_{latent}$ are situated in regions exhibiting the greatest degree of plastic flow. This indicates that the acceleration of chemistry in hotspots and shear bands is fundamentally connected to deviatoric material response through mechanochemistry and is not simply a product of local variations in pressure or density, which are largely uniform across the hotspot.

Figure 3 depicts the underlying distributions of $\tau_{rxn}$ times for the Clusters 2, 3, and 5. In all cases, we observe the expected Maxwell-Boltzmann-like distributions, but the effect of $U_{latent}$ is quite remarkable. Cluster 3 has a higher temperature than Cluster 2 and approximately the same $U_{latent}$, and as expected the reaction times are shorter. On the other hand, Cluster 5 has a similar temperature to Cluster 2, but with substantially higher $U_{Latent}$. This leads to almost the same effect as increasing temperature. While temperature and intramolecular strain energy are not uncorrelated (Pearson Correlation Coefficient: $0.46 \pm 0.04$; 95% confidence interval from bootstrapping), these distributions corroborate the trends in the cluster centroids that the $U_{Latent}$ of the molecule is directly affecting the reaction timescales. Cluster 3 and Cluster 5 have nearly the same distribution, despite being tails in the two different descriptors showing that $U_{Latent}$ is generally acting in a similar manner to the temperature, in terms of altering the overall kinetics and its statistics.

We note that the clustering analysis is based on an instantaneous measurement of molecular temperature and $U_{Latent}$ just before $t_1$, and these quantities vary during the time interval measured as $\tau_{Rxn}$. Specifically, the relaxation of $U_{4-body}$ with time shown in Figure 1 largely coincides with reactions that occur during the time interval between the first reaction at $t_1$ and proceeding up through ring breakage at time $t_2$. The decrease of $\tau_{Rxn}$, or more explicitly, the earlier C-C ring bond scission with respect to the first reaction, shows an inherent memory effect of the strain and temperature state at the point of first reaction. That is, after the strained bonds begin to break and reduce $U_{4-body}$, the downstream reactions are still affected by the energy state of the molecule at first reaction.

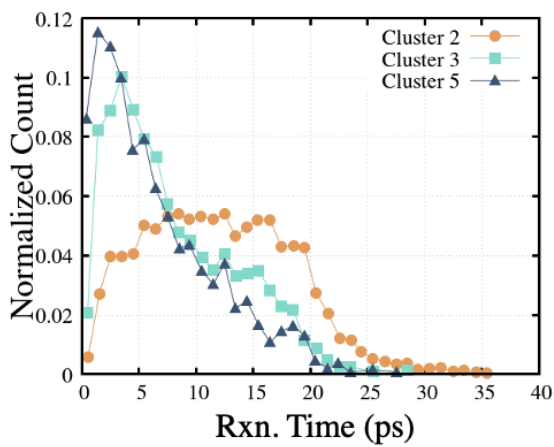

*Figure 3: Distributions for $\tau_{Rxn}$ for Clusters 2, 3, and 5, which represent the hotspot average, high temperature tail, and high intramolecular strain tail, respectively.*

Having established that multiple internal state variables of the decomposing molecule affect reaction kinetics, we now consider possible effects on decomposition paths. We focus on the first step of the decomposition of each molecule. Wu and Fried[48] showed that intramolecular hydrogen transfer is the most likely first reaction for TATB. By comparison, intermolecular H transfer has an ~10 kcal/mol higher activation barrier than the uni-molecular reaction and $NO_2$ scission exhibits an even higher barrier than both H-transfers. Enhanced $NO_2$ scission has also been identified as a key difference for reactions in TATB shear bands compared to the bulk crystal[17]. Supplemental Materials section SM-2 shows schematics of these possible reactions. Here, chemical bonds are determined by ReaxFF bond tables with a 0.5 threshold[46].

By assessing the number of unimolecular and bimolecular hydrogen transfers, $NO_2$ scission, and ring fracture 1st step reactions in each cluster, we can track trends with increasing temperature and intramolecular strain energy, shown in Figure 4a and 4b. Methods used to determine reaction paths are described in Supplemental Materials section SM-3. Clusters 1, 2, and 3 show little to no intramolecular strain energy, with Cluster 3 being a high temperature, low strain tail. For the most part, these clusters show almost all common reactions, mainly the intramolecular hydrogen transfer, with Cluster 3 having about 25% intermolecular hydrogen transfer. Clusters 4 and 5, which have increasing $U_{Latent}$ at similar average temperatures, show a few key trends. One is a large increase in the number of intermolecular hydrogen transfers, which ultimately make this route the dominant pathway, the other being an increase in the number of $NO_2$ scission reactions and a ring fracture events as the first step, albeit in trace amounts. (Note that this does not necessarily imply that these reactions are not common as downstream processes.) Cluster 6, which is a high temperature, mid-strain state, shows similar results to Cluster 5 with slightly lower counts for the "uncommon" first-step reactions. These agree well with previously reported premeditated TATB mechanochemistry[18].

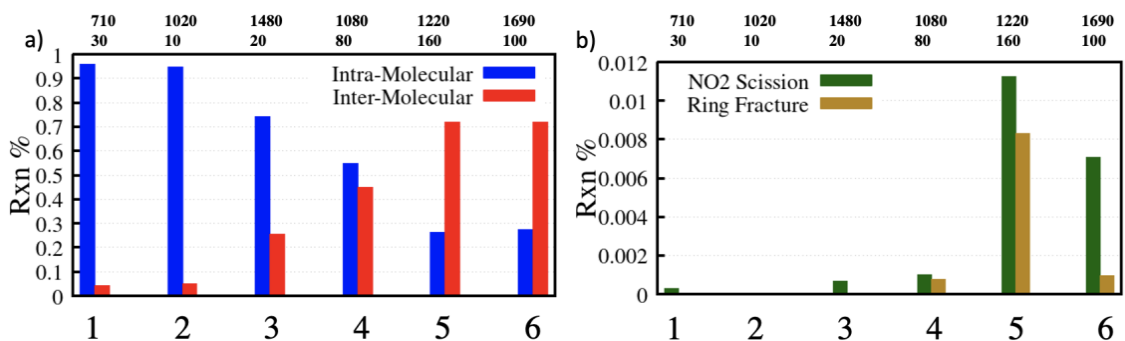

*Figure 4: Bar charts of the percentages of first step reaction pathways for 4 reaction classes, organized by the cluster IDs presented in Figure 2 and split into groups of common (a) and uncommon (b) reactions.*

Overall, we show through large-scale reactive MD simulations a significant acceleration and alteration of reactions in explosive hotspots due to local intramolecular strains that is not accounted for in "traditional" thermal chemistry models. This implies that HE shock initiation is fundamentally an extemporaneous mechanochemical process. A cluster analysis of the reacting molecules in our simulations links an acceleration of chemistry to increases in molecular temperature and $U_{Latent}$ in which the mechanochemistry effect is localized to areas that undergo significant plastic flow. Additionally, $U_{Latent}$ is found to alter the $\tau_{Rxn}$ distribution shape similar to temperature. An assessment of the first step reaction pathways for each cluster shows that increasing both temperature and $U_{Latent}$ can lead to common, alternate reaction paths becoming dominant, and uncommon reaction paths appear in the high temperature and high $U_{Latent}$ tails.

Our results show a direct relation between the level of intramolecular strain energy and reaction timescales, brought on in an extemporaneous manner. While these events occur at the level of material microstructure, they involve a wide range of subscale molecular states and deformations that lead to a highly mixed, ensemble response. Upscaling these physics holds promise to improve the transferability of continuum-level explosive models[17,42], and of mechanochemical effects in a wide range of other functional molecular and polymeric materials[31,32,34,35]. The connections between subscale molecular states and microstructural-level material dynamics revealed here indicate that a hybrid modeling approach may be best for rapidly screening for mechanochemical effects in other systems. This approach would combine large, yet efficient non-reactive simulations that explicitly resolve microstructural-level processes in atomic detail[43,44] in tandem with small targeted reactive simulations[18] that match distributions of $U_{Latent}$ states to accelerate the exploration chemical phase space governing extemporaneous mechanochemical events.

## Methods

Reactive, all-atom MD simulations were conducted using the ReaxFF forcefield[49] and the LAMMPS code[50]. We use the so-called ReaxFF 2018[51] parametrization, which has been used in past for TATB[52,53] and is known to correctly predict first step reaction pathways[48,53] and Hugoniot curves.[52] Trajectories were integrated using a 0.1 fs time step. Partial atomic charges were calculated using the charge equilibration scheme[54] at each time step with an accuracy threshold set to $1 \times 10^{-6}$.

The simulation cell was built using the triclinic $P\bar{1}$ crystal structure[55] with equilibrium lattice parameters determined with ReaxFF 2018 at (300 K, 1 atm). We used an orthorhombic representation of the triclinic cell previously reported in Ref. [52] that was obtained using the generalized crystal cutting method[56]. The orthorhombic cell was the replicated to be ~250 nm in the shock direction (**C** vector) and approximately 120 nm along cell vector **A**. The simulation cell is periodic in **A** and **B** and open in the shock direction. A cylindrical pore with a circular cross-section 40 nm in diameter was cut in the **A-C** plane at fractional coordinate (1/2, 1/3), yielding a system with a total of 8.6 million atoms. The system was thermalized over 50ps of isothermal-isochoric (NVT) dynamics at 300 K with a Langevin thermostat[57], re-initializing velocities following a Maxwell-Boltzmann distribution every 5 ps. The shock simulation was run using a reverse ballistic setup[58] with a particle velocity of 2.0 km/s with dynamics propagated using microcanonical (NVE) equations of motion.

Analysis of the trajectories was performed in a molecularly averaged, center of mass (COM) framework. Two primary molecular properties were considered: roto-vibrational kinetic energy and intramolecular strain energy. We refer to the first term as temperature and use units of kelvin for simplicity (using the classical specific heat of $\frac{69}{2}k_b$). The second term was defined above in Equation 1.

We defined a molecular, characteristic reaction timescale, $\tau_{Rxn} = t_2 - t_1$, based on two well-defined events in the molecule's reaction history. Bond scission was determined using ReaxFF bond orders. Time $t_1$ was defined as the first bond scission in the molecule. Time $t_2$ was defined as the first C-C (ring) bond scission. SM-4 shows snapshots of both $t_1$ and $t_2$ events. MD snapshots were saved and analyzed every 0.1 ps, and a bond break must be observed for three consecutive frames for it to be considered a reaction. We performed the same analysis for thermal decomposition simulations to obtain a benchmark for interpreting changes in energy, see SM-5.

# Acknowledgements

This work was supported by the Laboratory Directed Research and Development Program at Lawrence Livermore National Laboratory, project 18-SI-004 with Lara Leininger as P.I. Partial support was received from the US Office of Naval Research, Multidisciplinary University Research Initiatives (MURI) Program, Contract: N00014-16-1-2557. Program managers: Chad Stoltz and Kenny Lipkowitz. Partial support was received from the Army Research Laboratory and was accomplished under Cooperative Agreement Number W911NF-20-2-0189. Simulations were made possible by computing time granted to MPK through the LLNL Computing Grand Challenge, which is gratefully acknowledged. This work was performed under the auspices of the U.S. Department of Energy by Lawrence Livermore National Laboratory under Contract DE-AC52-07NA27344. It has been approved for unlimited release under document number LLNL-JRNL-834510-DRAFT.

# SM-1: Cluster Centroids

*SM Table 1: Cluster centroid data, as presented in Figure 2.*

| Cluster ID | Temperature (K) | $U_{Latent}$ (kcal/mol) | $\tau_{Rxn}$ (ps) |
|---|---|---|---|
| 1 | 710 | 30 | 13.8 |
| 2 | 1020 | 10 | 11.9 |
| 3 | 1480 | 20 | 7.8 |
| 4 | 1080 | 80 | 8.9 |
| 5 | 1220 | 160 | 6.5 |
| 6 | 1690 | 100 | 4.8 |

# SM-2: Reaction Paths

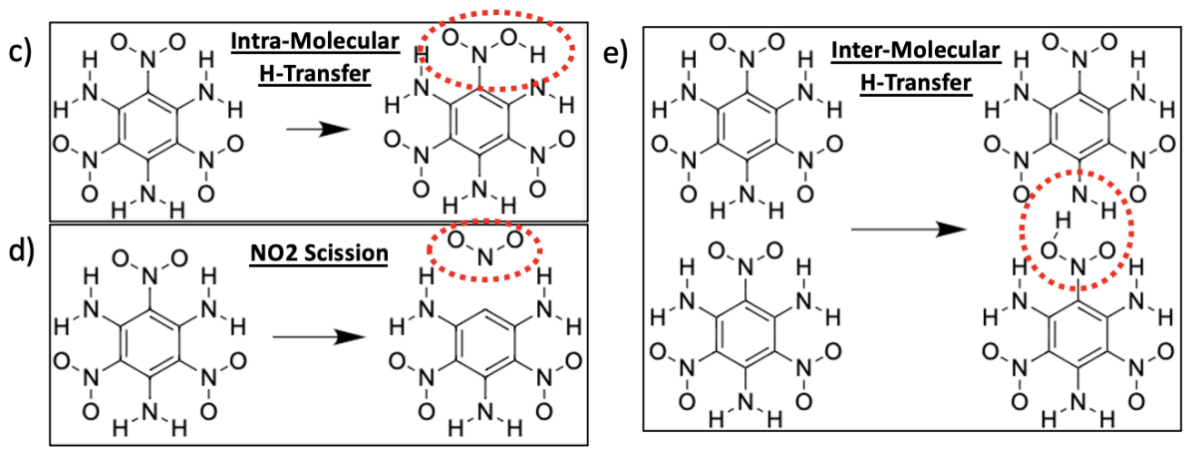

*SM- Figure 3: Schematics of the primary 1st step reaction paths for TATB.*

## SM-3: Chemical analysis details

ReaxFF simulations determine bonding on the fly via partial bond orders. From the bonding outputs, one can count the total number of bonds formed in a molecule, or with the atoms from an original molecule. For the 4 reactions tracked in this work, a distinction of reactions can be made by simply counting the bonds present. For each cluster, only trace amounts (1-2 molecules total) of reactions did not belong to one of the four reactions studied here.

We define as a molecular descriptor $D$ the total bond counts $B_i$ to atoms of each element type, which we track as a 4-dimentional vector $D = [B_C, B_H, B_N, B_O]$ for each molecule at each step throughout time. For instance, a C-O bond contributes +1 to $B_C$ and +1 to $B_O$. This analysis only considers atom-atom connections and does not treat double and triple bonds differently than single bonds.

An unreacted TATB molecule has bonding [18, 6, 18, 6]. When a molecule does not match this pattern for 3 consecutive frames (0.3 ps), we take first frame of the three to be time $t_1$.

Following $t_1$, the patterns for identifying each reaction are as follows:

Intra-molecular hydrogen transfer: [18, 6, 18, <8]

Inter-molecular hydrogen transfer: [18, 6, <18, 6]

NO$_2$ scission: [<18, 6, <18, 6]

Ring Fracture: [<18, 6, 18, 6]

## SM-4: Example molecular structures at characteristic times

Example structures at times between $t_1$ and $t_2$ for an intra-molecular hydrogen transfer reaction.

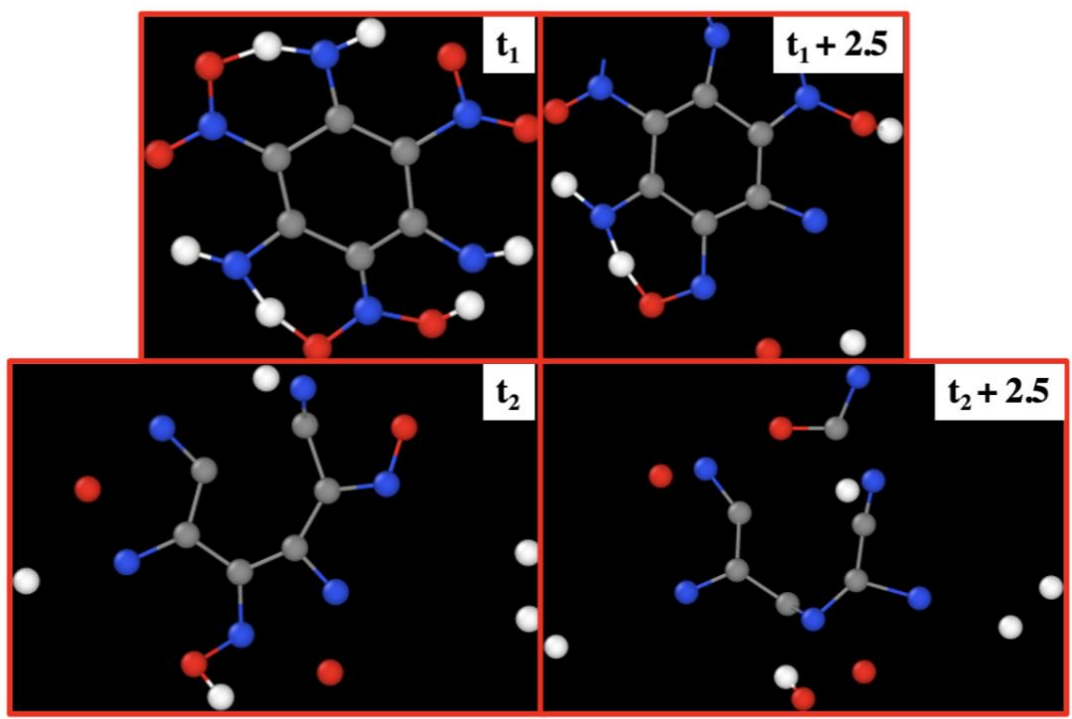

*SM Figure 1: Structures of a TATB molecule going through an intra-molecular hydrogen transfer reaction. Snapshots show an example molecule at the two characteristic times $t_1$ and $t_2$ defined in the main text, and at 2.5 ps after these times. Atoms are colored white, grey, blue, and red for H, C, N, and O.*

## SM-5: Time evolution of reactive events under isothermal conditions

We performed an isochoric-isothermal (NVT) decomposition simulation at 2500 K and ambient density to illustrate how $t_1$ and $t_2$ events evolve during the reaction as the total potential energy (PE) decays. The approach of cumulative $t_1$ events to unity corresponds to the completion of the 1st endothermic reaction step, and the approach of $t_2$ events to unity corresponds well to full exothermicity of the reaction.

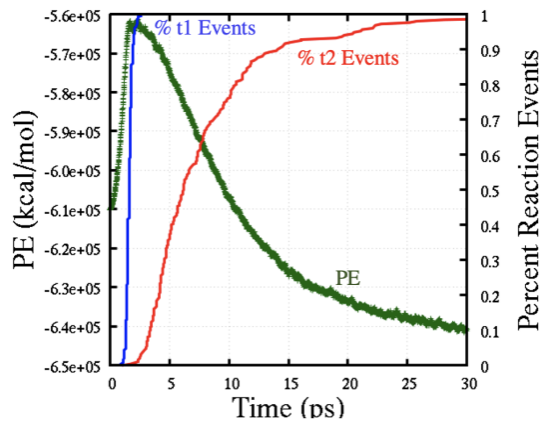

SM Figure 2: PE time history and reaction events time history. Total system PE (green) corresponds to the left y axis, both reaction events (red and blue) correspond to the right y axis.